\documentstyle[12pt]{article}

\def\be{\begin{equation}}
\def\ee{\end{equation}}
\def\bea{\begin{eqnarray}}
\def\eea{\end{eqnarray}}
\newcommand{\pa}{\partial^\mu}

\begin{document}
\begin{flushright}
\end{flushright}
\pagestyle{plain}
\begin{center}
\LARGE{\bf Geometrization of the Quantum Effects\\} \vspace{.5cm}
\small \vspace{1.5cm} {\Large{\bf H.
Motavali$^{1,2}$}}\footnote{e-mail address:
motavalli@tabrizu.ac.ir},~~ {\Large{\bf M.
Golshani$^{2}$}}\footnote{e-mail address: golshani@ihcs.ac.ir}
\\  \vspace{0.5cm}
\small{$^{1}$ Department of Theoretical Physics and Astrophysics,
Physics Faculty, Tabriz University,
PO Box 51664, Tabriz, Iran \\
$^{2}$ Institute for Studies in Theoretical Physics
and Mathematics, 19395-1795, Tehran, Iran}\\
\today
\end{center}
\vspace{.5cm} \small
\begin{abstract}
We present a conformally invariant generalized form of the free
particle action by connecting the wave and particle aspects
through gravity. Conformal invariance breaking is introduced by
choosing a particular configuration of dynamical variables. This
leads to the geometrization of the quantum aspects of matter.
\end{abstract}
\section{Introduction}
The general point of view of wave-particle duality proposed by de
Broglie ${}^1$ considers all atomic objects such as photons,
electrons, protons, etc. as consisting of the physical
association of two entities: (a) a wave, devoid of energy and
momentum but nevertheless objectively real and propagating in
space-time; (b) a particle, as a single tiny region of highly
localized energy incorporated in the wave, like a singularity in
motion ${}^2$. In fact corpuscle serves as carrier of the energy
belonging to the wave, while the wave constitutes an extended
entity surrounding the particle. Generally, wave and particle are
considered as different manifestations of a single system. In
particular, it is claimed they are two phenomena of essentially
different nature which constitutes the complementary aspects of a
single system.

The implementation of wave-particle duality in the causal
interpretation of quantum mechanic ${}^3$ leads to a distinct
feature, because one finds a direct interplay between particle
properties and those of waves through the quantum potential. This
would imply a substantial connection between wave and particle in
a typical dynamical process in quantum mechanics. In particular
one can infer the existence of the wave from the observation on
the associated particle. This paper undertakes a preliminary step
towards connecting this distinct face of wave-particle duality
with conformal invariance. We shall study the conformal invariant
coupling of a relativistic particle action to a scalar field
through the gravity. For the resulting theory, we identify two
different conformal frames. In the first frame, which we call
classical, the usual particle properties are established. We
shall then define a second frame, which we call quantum, in terms
of a local change of the classical frame. In this frame we
establish a dynamical interplay between a particle and the
applied conformal factor. This suggests that the association of a
wave with a particle in the causal interpretation of quantum
mechanics may be an irreducible effect of a conformal invariant
coupling of the particle to gravity. We shall use units in which
$\hbar=c=1$.

\section {The Conformal invariant coupling of a Particle to Gravity}
We consider a free particle action functional consisting of a
real scalar field $\phi$, as follows
\begin{eqnarray}
S_p[\phi]=-\frac{1}{2}{ \int d^4x {\sqrt {-g}}( {\phi}^2
g_{\mu\nu} \pa S \partial^\nu S + \frac{\lambda}{2} \phi^4 ) }
\end{eqnarray}
where $g_{\mu \nu}$ is the Riemannian metric, S is a
Hamilton-Jacobi function associated with the motion of free
particle and $\lambda$ is a dimensionless
parameter which depends on the particle properties.\\
It must be noted that, the action (1) dose not have the kinetic
energy of the scalar field $\phi$. In order to complete the
action (1), one may consider the kinetic term as an action
functional which contains the real scalar field $\phi$ and the
gravitational field in the form
\begin{eqnarray}
S_w[\phi]=- \frac{1}{2}{ \int d^4x \sqrt{-g}  (g_{\mu \nu} \pa
\phi \partial^\nu \phi+ \frac{1}{6}R\phi^2)}.
\end{eqnarray}
Here R is the Ricci curvature associated with the Riemannian
metric \footnote{ Our sign convention is given by $R_{\mu \nu}
\sim \Gamma_{\mu \nu, \alpha}^{\alpha} -\Gamma_{\mu \alpha,
\nu}^{\alpha} $ and the signature is $(-,+++)$.}. Thus the total
action is obtained by taking (1) and (2) together, giving us
\begin{eqnarray}
S[\phi] &=&{ S_p[\phi]+S_w[\phi]}   \\     \nonumber
       &=&{ - \frac{1}{2}{ \int  d^4x  \sqrt{-g} \{ g_{\mu \nu} \pa \phi \partial^\nu \phi
        +( g_{\mu \nu} \pa S \partial^\nu S + \frac{1}{6} R) {\phi}^2
       + \frac{\lambda}{2} \phi^4\} }}.
\end{eqnarray}

Variations of $S[\phi]$ with respect to $\phi$ gives
\begin{eqnarray}
\Box^g \phi - (g_{\mu \nu} \pa S \partial^\nu S +\frac{1}{6}R
)\phi- \lambda \phi^3  =0 \nonumber
\end{eqnarray}
or equivalently
\begin{eqnarray}
g_{\mu \nu} \pa S \partial^\nu S=-M^2 +\frac{ \Box^g
\phi}{\phi}-\frac{1}{6}R
\end{eqnarray}
where $M=(\lambda \phi^2)^{1/2}$ represents a varying
mass-parameter, and $\Box^g$ denotes the d'Alembertian associated
with the Riemannian metric $g_{\mu \nu}$, namely
\begin{eqnarray}
\Box^g=(-g)^{- \frac{1}{2}} \partial_\mu [(-g)^{ \frac{1}{2}} g^{
\mu \nu} \partial_\nu]. \nonumber
\end{eqnarray}
We call Eq. (4), the generalized Hamilton-Jacobi equation for a
particle, in which the particle's S-function is dynamically
coupled to gravity and the scalar field $\phi$.

Variations of $S[\phi]$ with respect to $g_{ \mu \nu}$ and $S$
leads to
\begin{eqnarray}
\phi^2 G_{\mu \nu}+(g_{\mu \nu} \Box^g - D_\mu D_\nu)\phi^2 +
6(D_\mu \phi D_\nu \phi+ \phi^2 D_\mu S D_\nu S)  \\ \nonumber
-3g_{\mu \nu}(D_\alpha \phi D^\alpha \phi + \phi^2 D_\alpha S
D^\alpha S)- \frac{3}{2} \lambda \phi^4 g_{\mu \nu}  =0
\end{eqnarray}
and
\begin{eqnarray}
\partial_\mu (\sqrt {-g} \phi^2 g^{ \mu \nu} \partial_\nu S)=0
\end{eqnarray}
where in Eq. (5), $G_{\mu \nu}=R_{\mu \nu}-\frac{1}{2}g_{\mu
\nu}R$~~ is the Einstein tensor, and $D_\mu$ represents the
covariant derivative. One may notice that the dynamical coupling
of the particle's S-function to gravity is entirely encoded in
the Eqs. (5) and (6) only. In fact the generalized
Hamilton-Jacobi equation (4) contains no new informations.
Actually, we may take the trace of Eq. (5) to obtain
\begin{eqnarray}
\phi \{ \Box^g \phi - (g_{\mu \nu} \pa S \partial^\nu S
+\frac{1}{6}R )\phi- \lambda \phi^3 \} =0 \nonumber
\end{eqnarray}
which, leads directly to the Eq. (4). This feature is a
consequence of the fact that the total action (3) is exactly
invariant under the  conformal transformation
\begin{eqnarray}
\left\{ \matrix{
g_{\mu \nu} \longrightarrow \Omega^2(x) g_{\mu\nu} \nonumber\\
\phi \longrightarrow \Omega^{-1}(x) \phi}.\right.
\end{eqnarray}
This implies that the theory introduced by the action (3) can be
considered in various (conformal) frames depending on the
particular choice of the local standard of length ${}^{4,5}$. A
simple conformal frame may be characterized by the condition
$\phi=\phi_{0}$= const. We may call it the classical frame. In
this frame Eqs. (5) and (6) take, respectively, the forms
\begin{eqnarray}
 G_{\mu \nu}+6D_\mu S D_\nu S-3g_{\mu \nu} D_\alpha S D^\alpha S-
\frac{3}{2} \lambda \phi_0^2 g_{\mu \nu}  =0
\end{eqnarray}
and
\begin{eqnarray}
\Box^g S=0.
\end{eqnarray}
These equations contain the entire dynamical coupling of the
particle's S-function to gravity in the classical frame. In this
frame we find, by taking the trace of Eq. (8), the generalized
Hamilton-Jacobi equation
\begin{eqnarray}
g_{\mu \nu} \pa S \partial^\nu S=-M_0^2 -\frac{1}{6}R
\end{eqnarray}
with $M_0 =(\lambda \phi_0^2)^{1/2}$ being a constant
mass-parameter.

A varying configuration of the scalar field $\phi$ relative to
the constant value $\phi=\phi_0$ can be chosen as a
representation of a new frame which we may call the quantum
frame. Such a varying configuration can represent fluctuations of
the scalar field $\phi$ around the constant value $\phi_0$. These
fluctuations may be described in terms of a local change of the
classical frame. To be specific, one may define a quantum frame
by applying a conformal transformation (7)  to the classical
frame with
\begin{eqnarray}
\Omega(x)=\frac{\phi_0}{\phi}. \nonumber
\end{eqnarray}
Using the transformation law of the Ricci tensor $R_{\mu\nu}$
under (7), we then find ${}^6$
\begin{eqnarray}
R_{\mu\nu} \longrightarrow R_{\mu\nu}+ \Omega^{-2} \{4 {D_\mu
\Omega D_\nu \Omega}-2\Omega{D_\mu D_\nu \Omega}
-g_{\mu\nu}g^{\alpha \beta}(\Omega{D_\alpha D_\beta
\Omega}+{D_\alpha \Omega D_\beta \Omega}) \} \nonumber
\end{eqnarray}
\begin{eqnarray}
G_{\mu\nu} \longrightarrow G_{\mu\nu}+\Omega^{-2} \{ (g_{\mu\nu}
\Box^g
 - D_\mu D_\nu)\Omega^2 + 6D_\mu \Omega D_\nu \Omega-3g_{\mu \nu} D_\alpha \Omega D^\alpha \Omega \}.
\nonumber
\end{eqnarray}
Consequently the Eq. (8) transforms as
\begin{eqnarray}
\Omega^2 G_{\mu \nu}+ (g_{\mu \nu} \Box^g - D_\mu D_\nu)\Omega^2
+ 6(D_\mu \Omega D_\nu \Omega+ \Omega^2 D_\mu S D_\nu S)  \\
\nonumber -3g_{\mu \nu}(D_\alpha \Omega D^\alpha \Omega +
\Omega^2 D_\alpha S D^\alpha S)- \frac{3}{2} M_0^2 \Omega^2
g_{\mu \nu} =0. \nonumber
\end{eqnarray}

Taking the trace of Eq. (11), we find
\begin{eqnarray}
g_{\mu \nu} \pa S \partial^\nu S=-{M_0}^2 + \frac{ \Box^g
\Omega}{\Omega}-\frac{1}{6}R.
\end{eqnarray}

Comparing Eqs. (10) and (12), we find that, in the quantum frame
the Hamilton-Jacobi equation is identical with the classical
frame, except for the particle mass which is modified by an extra
term $\frac{ \Box^g \Omega}{\Omega}$. The appearance of this term
is a consequence of the local change of the classical frame by
the conformal transformation (7). The interpretation of this
extra term is postponed to the next section.

It must be pointed out however, that Eq. (9) is left  unchanged
in the quantum frame. It is convenient to rewrite it in the
following form for later usage
\begin{eqnarray}
\partial_\mu (\sqrt {-g} \Omega^2 g^{ \mu \nu} \partial_\nu S)=2\sqrt{-g}{\cal{M}} \Omega^2 \frac{d}{d \tau} \ln \Omega
\end{eqnarray}
where we have used
\begin{eqnarray}
\partial_\mu S={\cal{M}} U_\mu  ~~~~~~~~~and~~~~~~~~~~   U_\mu \partial^\mu =\frac{d}{d \tau}
\nonumber
\end{eqnarray}
in which $\tau$ is a parameter along the particle trajectory and
the value ${\cal {M}}$ is dependent on the frame one is using.
\section{Derivation of a pilot wave }
In the quantum frame, it is possible to assign a pilot wave to
the particle motion described by the Hamilton-Jacobi equation
(12). Consider the wave
\begin{eqnarray}
\psi= \Omega~ e^{iS} \nonumber
\end{eqnarray}
which satisfies, as a consequence of Eqs. (12) and (13), the
equation
\begin{eqnarray}
\Box^g \psi- {M_0}^2 \psi=(\frac{1}{6}R+2i {\cal{M}} \frac{d}{d
\tau} \ln |\psi|)\psi. \nonumber
\end{eqnarray}

In the background approximation $g_{\mu\nu} \longrightarrow
\eta_{\mu \nu}$, where $\eta_{\mu\nu}$ is the Minkowski metric,
and in the limit in which any conceivable dependence of the
conformal factor $\Omega$ on the parameter $\tau$ along the
particle trajectory is ignored, this equation reduces to the
massive Klein-Gordon equation
\begin{eqnarray}
\Box \psi- {M_0}^2 \psi= 0      \nonumber
\end{eqnarray}
where, $\Box$ denotes the d'Alembertian associated with the
Minkowski-metric. The merit of introducing the wave $\psi$ is
that it acts as a sort of a pilot wave in the sense of causal
interpretation of quantum mechanics ${}^{3,7}$, the term
$\frac{\Box^g \Omega}{\Omega} $ on the right hand side of Eq. (12)
being the associated quantum potential.

\end{document}